\documentclass{elsart}
\usepackage{natbib}
\usepackage{amssymb}

\begin{document}

\begin{frontmatter}



\title{Relativistic generalization of Brownian Motion}


\author{T.~Koide and T.~Kodama}

\address{Instituto de F\'{\i}sica, Universidade Federal do Rio de Janeiro, C. P.
68528, 21945-970, Rio de Janeiro, Brazil}

\begin{abstract}
The relativistic generalization of the Brownian motion is discussed. 
We show that the transformation property of the noise term is determined by requiring for the equilibrium distribution function to be Lorentz invariant, such as the J\"uttner distribution function.
It is shown that this requirement generates an entanglement between the force term and the noise so that the noise itself should not be a covariant quantity.
\end{abstract}


\end{frontmatter}


\section{Introduction}

\label{}

Presently the relativistic hydrodynamics is found to be one of the important tool to describe the collective behaviors of relativistic heavy-ion collisions \cite{QM}. However, there are several problems yet to be clarified when one wants to extract quantitative information of the QGP formed in the collision \cite{Kodama}. 
Some of them are associated with the difficulties of obtaining physical inputs such as equation of states, initial conditions or hadronization processes.
Other questions are associated with the basic principle of hydrodynamics itself, that is, the validity of the assumption of the local thermal equilibrium. 
Thus in order to extract conclusive information on the dynamics of relativistic heavy-ion collisions, we need to study these questions from
various point of view. For example, to investigate more detailed behaviors
than the simplest ideal fluid models and sudden freezeout mechanism, we need
to develop mesoscopic theories like the Boltzmann equation, the Langevin
equation and so on \cite{Nonaka}, to connect the hydrodynamic model to real
observables. 

The relativistic Boltzmann equation approach has been applied by several
authors to estimate the viscosity, heat conduction, etc \cite{Boltzmann}.
Strictly speaking, the predictions besed on the Boltzmann equation is
reliable only in the Boltzmann-Grad limit and its applicability seems to be
quite limited to extract the correct values of required observables in many
realistic situations.

As one possible alternative, we may consider the Langevin approach. This
approach has been studied in low energy heavy-ion physics\cite%
{HeavyIonlangevin} and also recently in relativistic cases particularly
related to the behavior of heavy quarks \cite{Langevin_heavyIon_Rel}. 
In the latter cases, the
relativistic covariance of the Langevin approach deserves a careful
attention. To the authors' knowledge, there are a very few studies on this
aspect of the Langevin equation \cite%
{Dud,Hak,Ben,Oron,Boyer1,Boyer2,Pos,Deb,Fra,Hanggi1,Hanggi2}, and the
conclusions achieved in the former studies seem to raise some basic
questions.

In this paper, we would like to discuss the relativistic generalization of
the Brownian motion. Actually the requirement of the relativistic covariance
casts non-trivial questions because of the following reasons. The first
point is the non-uniqueness of the discretization scheme of the Langevin
equation due to the nature of multiplicative noise of a covariant Langevin
equation for a relativistic particle. Even if the noise is additive in a
certain frame, it becomes multiplicative noise in other frame by the Lorentz
boost. As is well known (and we will see later), when we have a
multiplicative noise, the discretization scheme of the noise term is not
unique and the solution depends on the descretization scheme. That is, the
discretization scheme and the Lorentz transformation entangle each other.

The second point is the transformation property of the noise. In general,
the Langevin equation is composed by the force term and the noise term. In
most of works, the force term and the noise term are assumed to be independent Lorentz vectors, and thus transform independently. 
In the continuum limit, this could be a natural assumption as far as the time
evolution is defined in terms of the proper time of the particle. However,
on integration, we have to define the Stieltjes integral of the noise term
on a finite integration measure of the time. As a result, if the Lorentz
covariance is required for the noise term, then the equilibrium distribution
function depends on the integrating Lorentz frame and contradicts with the
scalar nature of the equilibrium distribution function.

In the following, we discuss the consistent relativistic generalization of
Brownian motion. In Sec. 2, we first discuss the Brownian motion of a
relativistic particle in the rest frame of a heat bath. We then analyze the
corresponding Fokker-Planck equation by introducing the several integration
schemes of the noise term. An explicit form of the distribution function at
thermal equilibrium is obtained. In Sec.3, we show that how the Langevin
equation obtained should transform under an arbitrary Lorentz boost of the
system. This is done by requiring that the equilibrium distribution should
be a scalar. The result show that the Langevin equation of the relativistic
Brownian motion has a non-trivial Lorentz transformation property. Summary
and concluding remarks are given in Sec. 4.

\section{Relativistic Brownian motion in the rest frame of heat bath}

We consider the Brownian motion of a relativistic particle with mass $m$ in
the 3+1 dimension. In the rest frame of a heat bath, we consider the
following Langevin equation (in the units of $c=1$), 
\begin{eqnarray}
\frac{d\mathbf{x}^{\ast }}{dt^{\ast }} &=&\frac{\mathbf{p}^{\ast }}{p^{0\ast
}},  \label{eqn:1} \\
\frac{d\mathbf{p}^{\ast }}{dt^{\ast }} &=&-\nu (p^{0\ast })\mathbf{p}^{\ast
}+\sqrt{2D(p^{0\ast })}\mathbf{F}(t^{\ast }),  \label{eqn:2}
\end{eqnarray}%
where $p^{0\ast }=\sqrt{(p^{\ast })^{2}+m^{2}}$. The parameter $\nu
(p^{0\ast })$ and $D(p^{0\ast })$ are assumed the Lorentz scalar functions
and characterizes the relaxation of the momentum and the strength of the
noise, respectively. Here the index ${^{\ast }}$ denotes the variables in
the rest frame of the heat bath. The Gaussian white noise $\mathbf{F}(t)$
has the following correlation properties, 
\begin{eqnarray}
\langle \mathbf{F}(t^{\ast })\rangle _{0} &=&0, \\
\langle \mathbf{F}^{i}(t^{\ast })\mathbf{F}^{j}(t^{\ast ^{\prime }})\rangle
_{0} &=&\delta _{ij}\delta (t^{\ast }-t^{\ast ^{\prime }}).
\end{eqnarray}%
The symbol $\left\langle X\right\rangle _{0}$ denotes the stochastic average
of $X$ in the rest frame of the heat bath (we refer to as RF).

Now we replace the Langevin equation with the stochastic differential
equation (SDE) 
\begin{eqnarray}
d\mathbf{x}^* &=& \frac{\mathbf{p}^*}{p^{0*}}dt^*, \\
d\mathbf{p}^* &=& - \nu(p^{0*})\mathbf{p}^* dt^* + \sqrt{2D(p^{0*})} d%
\mathbf{w}_{t^*}.  \label{eqn:SDE1}
\end{eqnarray}
Here we used 
\begin{eqnarray}
d\mathbf{w}_{t^*} \equiv \mathbf{w}_{t^* + dt^*} - \mathbf{w}_{t^*} = 
\mathbf{F}(t^*)dt^*,
\end{eqnarray}
with the help of the Wiener process $\mathbf{w}_{t^*}$. The correlations are
given by 
\begin{eqnarray}
\langle d\mathbf{w}^*_{t^*_i} \rangle_0 &=& 0 ,  \label{eqn:corr1} \\
\langle d\mathbf{w}^{i*}_{t^*_k}d\mathbf{w}^{j*}_{t^*_l} \rangle_0 &=& dt^*
\delta_{ij}\delta_{kl}.  \label{eqn:corr2}
\end{eqnarray}

The last term of Eq. (\ref{eqn:SDE1}) contains a kind of Stieltjes integral
of the stochastic variable. However, the definition of the required
Stieltjes integral is not unique. Here we consider the three typical cases.

\begin{enumerate}
\item Ito interpretation \cite{handbook}

In this case, the last term is interpreted as 
\begin{eqnarray}
\sqrt{2D(p^{0*})} d\mathbf{w}_{t^*} \longrightarrow [\sqrt{2D(p^{0*})} d%
\mathbf{w}^*_{t^*}]_{I} = \sqrt{2D(p^{0*}(t^*))} (\mathbf{w}^*_{t^* + dt^*}- 
\mathbf{w}^*_{t^*}).  \nonumber \\
\end{eqnarray}

\item Stratonovich-Fisk interpretation \cite{handbook}

In this case, the last term is interpreted as 
\begin{eqnarray}
\lefteqn{\sqrt{2D(p^{0*})} d\mathbf{w}_{t^*} \longrightarrow } &&  \nonumber
\\
&& [\sqrt{2D(p^{0*})} d\mathbf{w}^*_{t^*}]_{SF} = \frac{\sqrt{2D(p^{0*}(t^*
+ dt^*))}+\sqrt{2D(p^{0*}(t^*)})}{2} (\mathbf{w}^*_{t^* + dt^*}- \mathbf{w}%
^*_{t^*}).  \nonumber \\
\end{eqnarray}
By using the Ito formula (See appendix), we can show that the
Stratonovich-Fisk SDE is equivalent to the following Ito SDE, 
\begin{eqnarray}
d\mathbf{x}^* &=& \frac{\mathbf{p}^*}{p^{0*}} dt^*, \\
d\mathbf{p}^* &=& -\nu(p^{0*}) \mathbf{p}^* dt^* + \sqrt{D(p^{0*})}%
(\partial_{\mathbf{p}^*}\sqrt{D(p^{0*})}) dt^* + [\sqrt{2D(p^{0*})} d\mathbf{%
w}^*_{t^*}]_{I}.  \nonumber \\
\end{eqnarray}

\item H\"anggi-Klimontovich \cite{Hanggi3,Hanggi4,Kli}

In this case, the last term is interpreted as 
\begin{eqnarray}
\sqrt{2D(p^{0*})} d\mathbf{w}_{t^*} \longrightarrow [\sqrt{2D(p^{0*})}
dw^*_{t^*}]_{HK} = \sqrt{2D(p^{0*}(t^*+dt^*))} (\mathbf{w}^*_{t^* + dt^*}- 
\mathbf{w}^*_{t^*})  \nonumber \\
\end{eqnarray}
By using the Ito formula, we can show that the H\"anggi-Klimontovich SDE is
equivalent to the following Ito SDE, 
\begin{eqnarray}
d\mathbf{x}^* &=& \frac{\mathbf{p}^*}{p^{0*}} dt^*, \\
d\mathbf{p}^* &=& -\nu(p^{0*}) \mathbf{p}^* dt^* + 2\sqrt{D(p^{0*})}%
(\partial_{\mathbf{p}^*}\sqrt{D(p^{0*})}) dt^* + [\sqrt{2D(p^{0*})} d\mathbf{%
w}^*_{t^*}]_{I}.  \nonumber \\
\end{eqnarray}
\end{enumerate}

We are interested in the equilibrium distribution function described by
using these SDEs. For this purpose, we introduce the probability density 
\begin{equation}
\rho (\mathbf{x},\mathbf{p},t)=\langle \delta ^{(3)}(\mathbf{x}-\mathbf{x}%
^{\ast }(t^{\ast }))\delta ^{(3)}(\mathbf{p}-\mathbf{p}^{\ast }(t^{\ast
}))\rangle .  \label{rho}
\end{equation}%
Then, the time evolution of $\rho (\mathbf{x},\mathbf{p},t)$ is given by the
Fokker-Planck equation is 
\begin{eqnarray}
\partial _{t^{\ast }}\rho (\mathbf{x}^{\ast },\mathbf{p}^{\ast },t^{\ast })
&=& -\sum_{i}\partial _{\mathbf{x}^{\ast }}^{i}\frac{p^{i\ast }}{p^{0\ast }}%
\rho (\mathbf{x}^{\ast },\mathbf{p}^{\ast },t^{\ast })) +\sum_{i}\partial _{%
\mathbf{p}^{\ast }}^{i}(\nu (\mathbf{p}^{0\ast })p^{i\ast }\rho (\mathbf{x}%
^{\ast },\mathbf{p}^{\ast },t^{\ast }))  \nonumber \\
&& +\sum_{i}\partial _{\mathbf{p}^{\ast }}^{i}D^{1-\alpha }(\mathbf{p}%
^{0\ast })\partial _{\mathbf{p}^{\ast }}^{i}D^{\alpha }(\mathbf{p}^{0\ast
})\rho (\mathbf{x}^{\ast },\mathbf{p}^{\ast },t^{\ast }).
\end{eqnarray}
The parameter $\alpha $ denotes the different discritization scheme, $\alpha
=0,1/2$ and $1$ represents the H\"{a}nggi-Klimontovich, Stratonovich-Fisk
and Ito scheme, respectively.

The equilibrium distribution of the Fokker-Planck equation is 
\begin{equation}
\rho _{st}(\mathbf{p}^{\ast })\propto \exp {\ \left( -\int_{C}^{\mathbf{p}%
^{\ast }}d\mathbf{q \cdot }\frac{\mathbf{q }\nu (q^{0\ast })}{D(q^{0\ast })}%
-\alpha \ln D(p^{0\ast })\right) },  \label{eqn:st1}
\end{equation}%
where the path $C$ of the integral in the momentum space is arbitrary. 
This equilibrium distribution should be a Lorentz scalar by definition Eq.(\ref%
{rho}). As we will see later, we can define the transformation property of
the noise by using this fact.

\section{Relativistic Brownian motion in general frame}

We consider the reference frame which is moving with the velocity $\mathbf{V}
$ with respect to the rest frame of the heat bath (we refer to as simply 
\textit{MF-moving frame}). The four-momentum $dp^{\mu }$ in this frame is
then given by the Lorentz transformation of $dp^{\ast \mu }$ as%
\begin{eqnarray}
dp^{\mu } &=&\Lambda (\mathbf{V})dp^{\ast \mu }  \nonumber \\
&=&\left( 
\begin{array}{cc}
\gamma (\mathbf{V}) & \beta (V)\mathbf{n}^{T}\gamma (\mathbf{V}) \\ 
\beta (V)\mathbf{n}\gamma (\mathbf{V}) & \gamma (\mathbf{V})P_{\parallel
}+Q_{\perp }%
\end{array}%
\right) \left( 
\begin{array}{c}
dp^{0\ast } \\ 
d\mathbf{p}^{\ast }%
\end{array}%
\right) .
\end{eqnarray}%
Here, $P_{\parallel }=\mathbf{n}\mathbf{n}^{T}$ and $Q_{\perp
}=1-P_{\parallel }$ with $\mathbf{n}=\mathbf{V}/|\mathbf{V}|$. By using the
on mass-shell condition%
\[
dp^{0}=-\frac{\mathbf{p}}{p^{0}}\cdot d\mathbf{p} 
\]
we get the SDE in the MF as 
\begin{equation}
d\mathbf{p}^{i}=-\frac{\nu (u^{\mu }p_{\mu })\gamma (V)}{p^{0}}\left\{ p^{0}(%
\mathbf{p}^{i}-\beta (V)\mathbf{n}^{i}p^{0})+\beta (V)(p^{2}\mathbf{n}%
^{i}-p_{V}\mathbf{p}^{i})\right\} dt+[\mathbf{B}d\mathbf{w}_{t^{\ast
}}^{\ast }]^{i},  \label{eqn:SDEALT}
\end{equation}%
where $p_{V}=\mathbf{n}^{T}\mathbf{p}$ and 
\begin{eqnarray}
\mathbf{B}=\sqrt{2D(u^{\mu }p_{\mu })}\frac{\gamma ^{-1}(V)}{p^{0}-\beta
(V)p_{V}}\left\{ p^{0}P_{\parallel }+\gamma (V)\left( p^{0}-\beta
(V)p_{V}+\beta (V)(\mathbf{n}\cdot \mathbf{p}^{T})\right) Q_{\perp }\right\}
.  \nonumber \\
\end{eqnarray}
The last term in Eq.(\ref{eqn:SDEALT}) should be interpreted appropriately
according to the integration scheme, namely the Ito, Stratonovich-Fisk or H%
\"{a}nggi-Klimontovich cases.

The last term is not yet transformed because it contains the noise in RF $%
dw_{t^*}$. First of all, we assume that, even after the Lorentz boost, the stochastic part of the Brownian motion still preserves the property of the
Gaussain white noise, which is defined by 
\begin{eqnarray}
\langle d\mathbf{w}_{t}\rangle _{V} &=&0, \\
\langle d\mathbf{w}_{t_{l}}^{i}d\mathbf{w}_{t_{m}}^{j}\rangle _{V}
&=&dt\delta _{ij}\delta _{lm},
\end{eqnarray}%
Here the simbol $\left\langle X\right\rangle _{V}$ denotes the stochastic average of $X$ in the MF.

Then, at first sight, as was assumed in previous works \cite%
{Dud,Hak,Ben,Oron,Boyer1,Boyer2,Pos,Deb,Fra,Hanggi1,Hanggi2}, we may suppose
that the noise term is isolatedly Lorentz covariant and we could write 
\begin{eqnarray}
\langle d\mathbf{w}_{t}\rangle _{V} &=&\langle d\mathbf{w}_{t^{\ast }}^{\ast
}\rangle _{0}=0, \\
\tilde{\gamma}^{-1}(\mathbf{p})\langle d\mathbf{w}_{t_{k}}^{i}d\mathbf{w}%
_{t_{l}}^{j}\rangle _{V} &=&\gamma (\mathbf{p}^{\ast })\langle d\mathbf{w}%
_{t_{k}^{\ast }}^{i\ast }d\mathbf{w}_{t_{l}^{\ast }}^{j\ast }\rangle
_{0}=d\tau \delta _{ij}\delta _{kl},
\end{eqnarray}%
where 
\begin{eqnarray}
\tilde{\gamma}^{-1}(\mathbf{p})=(\Lambda (\mathbf{V})\Lambda (-\mathbf{p}%
^{\ast }))^{00}=\gamma (\mathbf{V})\gamma (\mathbf{p}^{\ast })(1-\beta (V)%
\frac{p_{V}^{\ast }}{p^{0\ast }}).
\end{eqnarray}
If this is the case, we could define the transformation property of the
noise as follows; 
\begin{eqnarray}
d\mathbf{w}_{t^{\ast }}^{\ast } &=&\sqrt{\frac{dt^{\ast }}{dt}}d\mathbf{w}%
_{t} = \sqrt{\gamma (\mathbf{p})\tilde{\gamma}^{-1}(\mathbf{p}^{\ast })}d%
\mathbf{w}_{t} =\gamma ^{1/2}(V)\sqrt{\frac{p^{0}-\beta (V)p_{V}}{p^{0}}}d%
\mathbf{w}_{t}.  \label{Trans1}
\end{eqnarray}%
However, as we will see in below, this transformation rule (\ref{Trans1}) does not give the correct properties of the Langevin equation. The
reason for this is that the Stieltjes integral associated with the noise
term is defined on the time interval $dt$, so that $d\mathbf{w}_{t^{\ast
}}^{\ast }$ is non-local in the time $t$. Thus the Lorentz transformation
entangles with the integration scheme in the order of $dt$. Then the noise
term itself is not covariant but constitutes a Lorentz vector together with
the force term. That is, the force part and the stochastic part can be
mixed. Then, the first order correlation calculated in the MF of the noise
term in the rest frame of the heat bath does not vanish, 
\begin{eqnarray}
\langle d\mathbf{w}_{t^{\ast }}^{\ast }\rangle _{V}\neq 0.
\end{eqnarray}
Thus we should consider, instead of Eq.(\ref{Trans1}) the following
transformation property of the noise, 
\begin{eqnarray}
d\mathbf{w}_{t^{\ast }}^{\ast } &=&\sqrt{\frac{dt^{\ast }}{dt}}d\mathbf{w}%
_{t}+\mathbf{C}_{\mathbf{p}}dt  \nonumber \\
&=&\gamma ^{1/2}(V)\sqrt{\frac{p^{0}-\beta (V)p_{V}}{p^{0}}}d\mathbf{w}_{t}+%
\mathbf{C}_{\mathbf{p}}dt.  \label{trans2}
\end{eqnarray}%
Here $\mathbf{C}_{\mathbf{p}}dt$ term acts as the force term in the MF,
which should be separated from the pure stochastic part $d\mathbf{w}_{t}$.
By using this definition and using the Ito formula, the Langevin equation is
given by 
\begin{eqnarray}
d\mathbf{p}^{i} &=&-\frac{\nu (u^{\mu }p_{\mu })}{p^{0}}\left\{ p^{0}(%
\mathbf{p}^{i}-\beta (V)\mathbf{n}^{i}p^{0})+\beta (V)(p^{2}\mathbf{n}%
^{i}-p_{V}\mathbf{p}^{i})\right\} dt  \nonumber \\
&&+(1-\alpha )\sum_{jk}\tilde{\mathbf{B}}^{jk}\partial _{\mathbf{p}}^{j}%
\tilde{\mathbf{B}}^{ik}dt+[\mathbf{B}\mathbf{C}_{p}]^{i}dt+[\tilde{\mathbf{B}%
}d\mathbf{w}_{t}]_{I}^{i},
\end{eqnarray}%
where 
\begin{eqnarray}
\tilde{\mathbf{B}}=\sqrt{\frac{\gamma (V)(p^{0}-\beta (V)p_{V})}{p^{0}}}%
\mathbf{B}
\end{eqnarray}
and the last term should be calculated according to the Ito scheme. It
should be noted that, to obtain a consistent result, the Ito formula to
convert the SDE from one scheme to the other and the replacement of the
noise by using Eq. (\ref{trans2}) must be used only after the transformation
of the noise, Eq. (\ref{trans2}).

Let us calculate the equilibrium distribution function in the MF. It is the
solution of the equation, 
\begin{eqnarray}
\left[ -\mathbf{A}^{i}-(1-\alpha )\sum_{jk}\tilde{\mathbf{B}}^{jk}\partial _{%
\mathbf{p}}^{j}\tilde{\mathbf{B}}^{ik}+\frac{1}{2}\partial _{\mathbf{p}}^{j}(%
\tilde{\mathbf{B}}\tilde{\mathbf{B}}^{T})^{ij}\right] \rho _{st}(\mathbf{p}%
)=0,
\end{eqnarray}
where 
\begin{eqnarray}
\mathbf{A}^{i}=-\frac{\nu (u^{\mu }p_{\mu })}{p^{0}}\left\{ p^{0}(\mathbf{p}%
^{i}-\beta (V)\mathbf{n}^{i}p^{0})+\beta (V)(p^{2}\mathbf{n}^{i}-p_{V}%
\mathbf{p}^{i})\right\} +[\mathbf{B}\mathbf{C}_{p}]^{i}.
\end{eqnarray}
The logarithmic derivative of the equilibrium distribution with respect to
the momentum is given by 
\begin{equation}
(\partial _{\mathbf{p}}^{i}\rho _{st})/\rho _{st}=2(\tilde{\mathbf{B}}\tilde{%
\mathbf{B}}^{T})_{ij}^{-1}\left( \mathbf{A}^{j}+(1-\alpha )(\partial _{%
\mathbf{p}}^{l}\tilde{\mathbf{B}}^{jk})\tilde{\mathbf{B}}^{lk}-\frac{1}{2}%
\partial _{\mathbf{p}}^{k}\left( \tilde{\mathbf{B}}\tilde{\mathbf{B}}%
^{T}\right) ^{jk}\right) .  \label{DeriMV}
\end{equation}%
If $\rho _{st}$ is a scalar function, it should coincide with the
logarithmic derivative of the same equilibrium distribution function
obtained in the RF. From Eq. (\ref{eqn:st1}), we have 
\begin{equation}
(\partial _{\mathbf{p}}^{i}\rho _{st})/\rho _{st}=\left( -\beta (V)\gamma (V)%
\frac{\mathbf{p\cdot n^{T}}}{p^{0}}+\gamma (V)P_{\parallel }+Q_{\perp
}\right) ^{ik}\left( \frac{\nu (p^{0\ast })}{D(p_{0\ast })}\mathbf{p}^{\ast
}+\alpha \frac{D^{\prime }(p^{0\ast })}{D(p^{0\ast })}\frac{\mathbf{p}^{\ast
}}{p^{0\ast }}\right) ^{k},  \label{DeriRF}
\end{equation}%
where $D^{\prime }(p^{0})=dD(p^{0})/dp^{0}$. To compare Eq.(\ref{DeriMV})
with Eq.(\ref{DeriRF}), it is convenient to decompose the vector $\mathbf{C}%
_{\mathbf{p}}$ into two directions, 
\begin{eqnarray}
\mathbf{C}_{\mathbf{p}}=\mathbf{C}_{\mathbf{p}}^{\perp }(\mathbf{p}-p_{V}%
\mathbf{n})+\mathbf{C}_{\mathbf{p}}^{\parallel }\mathbf{n}.
\end{eqnarray}
Then%
\begin{eqnarray}
\mathbf{C}_{\mathbf{p}}^{\perp } &=&-(1-\alpha )\sqrt{\frac{D(u^{\mu }p_{\mu
})}{2}}\frac{\beta (V)\gamma (V)}{(p^{0})^{2}}\frac{p_{V}-\beta (V)p^{0}}{%
p^{0}-\beta (V)p_{V}}, \\
\mathbf{C}_{\mathbf{p}}^{\parallel } &=&\alpha \sqrt{\frac{D(u^{\mu }p_{\mu
})}{2}}\frac{1}{(p^{0})^{2}}\frac{\beta (V)m^{2}}{p^{0}-\beta (V)p_{V}} 
\nonumber \\
&&+(1-2\alpha )^{2}\sqrt{\frac{D(u^{\mu }p_{\mu })}{2}}\frac{\beta (V)\gamma
(V)}{(p^{0})^{2}}(p^{0}-\beta (V)p_{V})(d-1)  \nonumber \\
&+&(1-\alpha )\sqrt{\frac{D(u^{\mu }p_{\mu })}{2}}\frac{\beta (V)\gamma (V)}{%
(p^{0})^{2}}\left\{ \left( \frac{p^{2}}{p^{0}}-\beta (V)p_{V}\right) -\left(
p^{0}p_{V}-\beta (V)p^{2}\right) \frac{1}{p^{0}}\frac{p_{V}-\beta (V)p^{0}}{%
p^{0}-\beta (V)p_{V}}\right\} ,  \nonumber \\
\end{eqnarray}%
where $d$ denotes the dimension of the space part. One can see that the form
of $\mathbf{C}_{\mathbf{p}}$ depends only on the parameter of the strength
of the noise $D(p^{0*})$, although the original Langevin equation contains
two parameters, $\nu(p^{0*})$ and $D(p^{0*})$. We further notice that the $%
\mathbf{C}_{\mathbf{p}}$ does not vanish by changing $D(p^{0*})$. That is,
in order to keep the scalar property of the equilibrium distribution
function, $\mathbf{C}_{p}$ cannot be null. It means that we cannot use the commonly assumed Lorentz covariant noise Eq.(\ref{Trans1}).

When the equilibrium distribution function is given by the J\"{u}ttner
function, 
\begin{eqnarray}
\rho _{st}=Const.\times e^{-\beta u^{\mu }p_{\mu }},
\end{eqnarray}
then the parameters of the SDE satisfy the following relation, 
\begin{equation}
\nu (u^{\mu }p_{\mu })=\frac{1}{u^{\mu }p_{\mu }}(\beta D(u^{\mu }p_{\mu
})-\alpha D^{\prime }(u^{\mu }p_{\mu })),  \label{Einstein}
\end{equation}%
which is the generalized Einstein's dissipation-fluctuation relation of the
relativistic Brownian motion.

\section{Summary and Concluding remarks}

In this work, we discussed the generalization of the Brownian motion of a
relativistic particle. First, we prepared a SDE in the rest frame of a heat
bath. We then derived the equilibrium distribution function by calculating
the corresponding Fokker-Planck equation. Next, by carrying out the Lorentz
boost explicitly requiring the on-mass-shell condition, we obtained the
boosted SDE in the moving frame with respect to the heat bath. The
covariance of the SDE requires that the noise is essentially multiplicative
in general frame. We also found that the Lorentz boost induces a non-trivial
entanglement between the force term and the noise term. We demonstrated that
the commonly used Lorentz invariant noise does not lead to an invariant
equilibrium distribution in the present formulation of the relativistic
Brownian motion.

The equation discussed here is different from those proposed in \cite%
{Deb,Hanggi1,Hanggi2}. In the RF, our equation is equivalent to their
equations. However, they assume the existence of the Lorentz covariant noise
term, and their equations in the MF are different from ours. To our best
knowledge, the consistency of the equilibrium distribution functions in the
RF and the MF is not discussed in their approaches.

The present study will be useful to investigate the Brownian motion of a
relativistic particle in a relativistically expanding heat bath such as jet
propagation in an expanding quark-gluon plasma formed in ultra-relativistic
heavy ion collisions. The further studies are in progress.

\ack{
The authors acknowledge stimulating discussions with G.S.Denicol. The
present work has been supported by CNPq, FAPERJ, CAPES and PRONEX. 
}


\appendix

\section{Ito formula}

\label{}

Let us consider an arbitrary function $f(\mathbf{x})$ and the evolution of $%
\mathbf{x}$ is given by the SDE, 
\begin{eqnarray}
d\mathbf{x}=\mathbf{A}dt+[\mathbf{B}d\mathbf{w}]_{I}.
\end{eqnarray}
Then, the variation of $f(\mathbf{x})$ is 
\begin{equation}
df(\mathbf{x})=\{\sum_{i}\mathbf{A}^{i}\partial _{i}f(\mathbf{x})+\frac{1}{2}%
\sum_{ij}[\mathbf{B}\mathbf{B}^{T}]^{ij}\partial _{i}\partial _{j}f(\mathbf{x%
})\}dt+\sum_{ij}[\mathbf{B}^{ij}\partial _{i}f(\mathbf{x})d\mathbf{w}%
^{j}]_{I}.  \label{Ito}
\end{equation}%
This is called the Ito formula \cite{handbook}. By using the Ito formula, we
obtain, 
\begin{eqnarray}
\lbrack \mathbf{B}d\mathbf{w}]_{SF} &=&[\mathbf{B}d\mathbf{w}]_{I}^{i}+\frac{%
1}{2}\sum_{jk}\mathbf{B}^{jk}\partial _{j}\mathbf{B}^{ik}dt, \\
\lbrack \mathbf{B}d\mathbf{w}]_{HK}^{i} &=&[\mathbf{B}d\mathbf{w}%
]_{I}^{i}+\sum_{jk}\mathbf{B}^{jk}\partial _{j}\mathbf{B}^{ik}dt.
\end{eqnarray}%
Thus we can conclude as follows. When we have the Stratonovich-Fisk SDE, 
\begin{eqnarray}
d\mathbf{x}=\mathbf{A}dt+[\mathbf{B}d\mathbf{w}]_{SF},
\end{eqnarray}
this is equivalent to the Ito SDE, 
\begin{eqnarray}
d\mathbf{x}^{i}=\{\mathbf{A}^{i}+\frac{1}{2}\sum_{jk}\mathbf{B}^{jk}\partial
_{j}\mathbf{B}^{ik}\}dt+[\mathbf{B}d\mathbf{w}]_{I}.
\end{eqnarray}
When we have the H\"{a}nggi-Klimontovich SDE, 
\begin{eqnarray}
d\mathbf{x}=\mathbf{A}dt+[\mathbf{B}d\mathbf{w}]_{HK},
\end{eqnarray}
this is equivalent to the Ito SDE 
\begin{eqnarray}
d\mathbf{x}^{i}=\{\mathbf{A}^{i}+\sum_{jk}\mathbf{B}^{jk}\partial _{j}%
\mathbf{B}^{ik}\}dt+[\mathbf{B}d\mathbf{w}]_{I}.
\end{eqnarray}


\end{document}